\documentclass[aps,prb,twocolumn,amssymb,superscriptaddress]{revtex4-1}

\usepackage{amsmath}
\usepackage{graphicx}
\usepackage{color}

\usepackage{bbding}
\usepackage{stmaryrd}
\usepackage{wasysym}

\usepackage{upgreek}
\usepackage{epsfig}
\usepackage{bm}
\usepackage{float}

\usepackage{url}
\usepackage{multirow}

\begin{document}

\title{Ground-state phase diagram of the frustrated spin-1/2 two-leg honeycomb ladder}
\author{Qiang Luo}
\affiliation{Department of Physics, Renmin University of China, Beijing 100872, China}
\author{Shijie Hu}
\email[]{shijiehu@gmail.com}
\affiliation{Department of Physics and Research Center Optimas, Technische Universit\"at Kaiserslautern, 67663 Kaiserslautern, Germany}
\author{Jize Zhao}
\affiliation{Center for Interdisciplinary Studies, Lanzhou University, Lanzhou 730000, China}
\author{Alexandros Metavitsiadis}
\affiliation{Department of Physics and Research Center Optimas, Technische Universit\"at Kaiserslautern, 67663 Kaiserslautern, Germany}
\affiliation{Institute for Theoretical Physics, Technical University Braunschweig, 38106 Braunschweig, Germany}
\author{Sebastian Eggert}
\affiliation{Department of Physics and Research Center Optimas, Technische Universit\"at Kaiserslautern, 67663 Kaiserslautern, Germany}
\author{Xiaoqun Wang}
\email[]{xiaoqunwang@sjtu.edu.cn}
\affiliation{Key Laboratory of Artificial Structures and Quantum Control (Ministry of Education), School of Physics and Astronomy, Tsung-Dao Lee Institute, Shanghai Jiao Tong University, Shanghai 200240, China}
\affiliation{Collaborative Innovation Center for Advanced Microstructures, Nanjing 210093, China}

\date{\today}

\begin{abstract}
%
%
We investigate a spin-$1/2$ two-leg honeycomb ladder with frustrating 
next-nearest-neighbor (NNN) coupling along the legs, which is equivalent to 
two $J_1$-$J_2$ spin chains coupled with $J_\perp$  at odd rungs.
The full parameter region of the model is systematically studied 
using conventional and infinite density-matrix renormalization group as well as bosonization.
The rich phase diagram consists of five distinct phases: 
a Haldane phase, a NNN-Haldane phase and a staggered dimer phase when $J_{\perp} < 0$;
a rung singlet phase and a columnar dimer phase when $J_{\perp} > 0$.
An interesting reentrant behavior from the dimerized phase into the Haldane phase 
is found as the frustration $J_2$ increases.
The universalities of the critical phase transitions are fully analyzed.
Phase transitions between dimerized and disordered phases belong to the two-dimensional 
Ising class with central charge $c=1/2$.
The transition from the Haldane phase to NNN-Haldane phase is of a weak 
topological first order,
while the continuous transition between the Haldane phase and rung singlet phase has 
central charge $c=2$.
\end{abstract}
%


\maketitle

\section{Introduction}
Frustration due to competing interactions in quantum spin systems 
often gives rise to interesting and rich
quantum phase diagrams.\cite{Diep_2005,Zhang_2016}
The prototypical example of a spin-1/2 chain with frustrating 
next-nearest-neighbor~(NNN) interactions $J_2$ is known to undergo a
{\it Berezinskii-Kosteriz-Thouless}~(BKT) transition from a gapless
{\it Tomonaga-Luttinger} liquid (TLL) phase to a gapped dimerized phase
with spontaneously broken translational symmetry\cite{Affleck_1990,Okamoto_1992,Castilla_1995,Eggert_1996,White_1996} at $J_{2,c} \simeq 0.241167J_1$ where $J_1$ 
is the nearest neighbor (NN) interaction along the chain.
At the {\it Majumdar-Ghosh} point\cite{Majumdar_1969_1,Majumdar_1969_2} $J_{2}=J_1/2$ 
the ground state is known exactly and for $J_{2}>J_1/2$
incommensurate spiral spin-spin correlations emerge.\cite{White_1996,Deschner_2013}
For comparison the spin-1 chain is in a gapped 
Haldane state\cite{Haldane_1983_1,Haldane_1983_2} with topological string 
order\cite{den_Nijs_1989,Kennedy_1992} for 
small $J_2$, but 
the correlations become incommensurate for $J_{2} \agt 0.284 J_1$
and a transition to a NNN-Haldane phase occurs at $J_{2,c} \simeq 0.7444J_1$,
where valance bonds link NNN sites,\cite{Kolezhuk_1996,Kolezhuk_1997,Hikihara_2000,Kolezhuk_2002,Pixley_2014,Chepiga_2016_1,Chepiga_2016_2,Chepiga_2016_3}
which is sometimes also referred to as a double Haldane phase.

Quantum spin ladders have been intensively studied in order to 
systematically extend the systems towards 
two-dimensions (2D).\cite{Dagotto_1992,Barnes_1993,White_1994,Azzouz_1994,Dagotto_1996,Schmidt_2003,Starykh_2004,Hung_2006,Kim_2008,Hikihara_2010,Barcza_2012,Vekua_2006,Liu_2008,Lavarelo_2011,Li_2012,Liu_2012,Sugimoto_2013,Metavitsiadis_2014,Metavitsiadis_2017} 
For unfrustrated two-leg spin-1/2 ladders the ground state is either in
a rung singlet~(RS) phase or
in a Haldane phase for positive or negative rung couplings $J_\perp$, respectively.
By including frustrating couplings between the chains $J_{\times}$
the possibility of a columnar dimerized~(CD) phase 
between the Haldane and RS phases was proposed,\cite{Xian_1995,Wang_2000,Starykh_2004} 
but the numerical evidence remains 
controversial for this case.\cite{Hung_2006,Kim_2008,Hikihara_2010,Barcza_2012}
On the other hand, small frustrating NNN couplings $J_2$ on the legs are 
known to induce and stabilize 
dimerized phases in the columnar or the staggered order,\cite{Vekua_2006,Liu_2008,Lavarelo_2011,Li_2012,Liu_2012,Sugimoto_2013,Metavitsiadis_2014,Metavitsiadis_2017} and in some cases 
reentrant behavior for larger $J_2$ may occur.\cite{Lavarelo_2011}

\begin{figure}[t]
\centering
\includegraphics[width=0.95\columnwidth, clip]{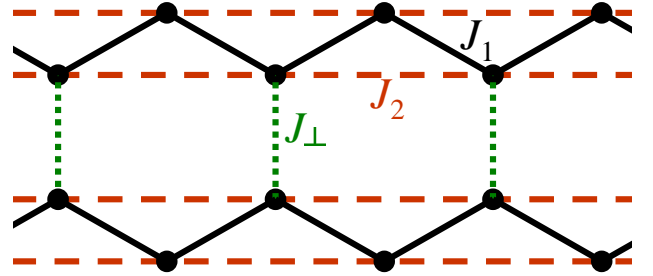}\\
\caption{Illustration of interactions on a frustrated two-leg spin-$1/2$ honeycomb ladder. }
\label{FIG_Model}
\end{figure}
We now analyze the honeycomb ladder as shown in Fig.~\ref{FIG_Model} consisting of 
two spin-1/2 legs with a rung-alternating coupling.\cite{Amiri_2015}
In this paper 
we focus on the effect of frustrating NNN couplings on the legs ($J_2$), while
frustrating inter-chain couplings ($J_\times$) may be considered in a later work.
As a 2D model the frustrated honeycomb model has received considerable attention
also in the context of longer range interactions and 
spin-liquid physics.\cite{Mattsson_1994,Singh_2011,Reuther_2011,Campbell_2012,Becca_2017}
Experimentally, the honeycomb ladder
has recently been realized in a four azide copper coordination compound 
[Cu$_{2}$L$^{1}$(N$_{3}$)$_{4}$]$_n$ (L$^{1}=2$, $6$-bis ($4$, $5$-dihydrooxazol-$2$-yl) pyridine), with 
antiferromagnetic intrachain coupling and 
ferromagnetic inter-chain coupling.\cite{Zhu_2013}  Using optical lattices a honeycomb 
structure can also be
constructed for interacting ultrcold gases.\cite{Esslinger_2018}
 
The paper is organized as follows:
In Sec.~\ref{SEC-Model}, we introduce the model Hamiltonian and discuss the symmetries.
The phase diagram from the numerical simulations is summarized and discussed in 
Sec.~\ref{SEC-PhaseDiagram}.
In Sec.~\ref{SEC-BONSONIZATION}, we use bosonization and the renormalization group (RG)
to obtain an analytic approach to predict the transition lines and compare with numerics 
before a detailed analysis of the critical behavior near the phase transitions is presented in Sec.~\ref{SEC-NatTrans}. 
The interesting behavior of the 
energy gaps is discussed in Sec.~\ref{SEC-SpinGap} and we conclude in 
Sec.~\ref{SEC-Con}.

\section{Model}\label{SEC-Model}
The frustrated honeycomb model shown in Fig.~\ref{FIG_Model} is described by the
spin-1/2 Hamiltonian
\begin{align}\label{HCHam}
\mathcal{H} = \mathcal{H}_{\rm legs} + \mathcal{H}_{\rm rungs}
\end{align}
with
\begin{equation}\label{HCHam-Leg}
\mathcal{H}_{\rm legs} = \sum_{\alpha=1,2} \sum_{j=1}^{L}  \left(J_1\mathbf{S}_{j,\alpha}\cdot\mathbf{S}_{j+1,\alpha} + J_2\mathbf{S}_{j,\alpha}\cdot
\mathbf{S}_{j+2,\alpha}\right)
\end{equation}
where $\alpha=1$, $2$ are the leg indices, and the interleg exchange on the odd rungs
reads
\begin{align}\label{HCHam-int}
\mathcal{H}_{\rm rungs} = &J_{\perp}\sum_{j=1}^{L}\frac{1-(-1)^j}{2}\mathbf{S}_{j,1}\cdot\mathbf{S}_{j,2}. 
\end{align}
%
Frustrating coupling ($J_\times$) between the legs can in principle also be 
included but will not be considered here.  
The leg couplings $J_1, \ J_2 > 0$ are both chosen to be 
antiferromagnetic, while the rung couplings $J_\perp$ can be 
ferromagnetic or antiferromagnetic.  For couplings on every rung the known instabilities 
are either the RS phase or the Haldane phase, depending on the 
sign,\cite{Dagotto_1992,Barnes_1993,White_1994,Azzouz_1994,Dagotto_1996,Schmidt_2003}
but the alternation may open the possibility for a richer phase diagram especially 
in connection with the frustration $J_2$, which tends to induce ordered dimerized 
phases.\cite{Vekua_2006,Liu_2008,Lavarelo_2011,Li_2012,Liu_2012,Sugimoto_2013,Metavitsiadis_2014,Metavitsiadis_2017}

Obviously, the total spin ${\mathbf{S}}_{t}=\sum_{j,\alpha} {\mathbf{S}}_{j,\alpha}$ 
commutes with model Hamiltonian, due to the global $SU(2)$ symmetry.
There is a mirror symmetry under exchange of the leg indices.
For periodic boundary conditions (PBC)
the translational symmetry by {\it two} rungs is maintained, but there is no 
parity symmetry along the legs.  Only for an odd number of rungs $L$ and open boundary conditions (OBC) such a parity 
symmetry exists with respect to the middle rung.

\section{Ground State Phase Diagram}\label{SEC-PhaseDiagram}
\begin{figure}[t]
\centering
\includegraphics[width=0.95\columnwidth, clip]{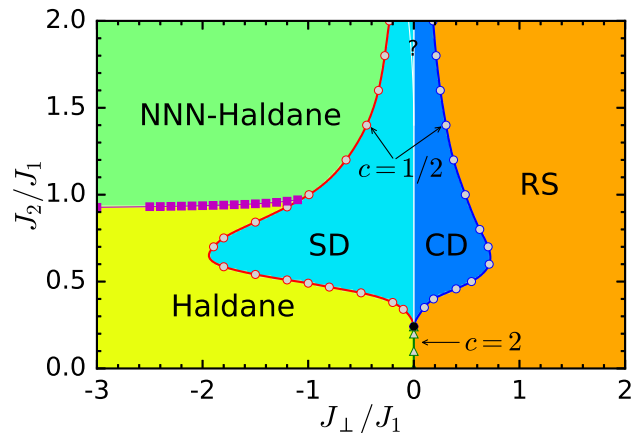}\\
\caption{Ground state phase diagram of the frustrated two-leg honeycomb ladder.
The ordered dimer phases SD and CD are separated by a first order transition
and Haldane to NNN-Haldane is also of weak first order.  All other phase transitions
are continuous ($\circ$: $c=1/2$; $\vartriangle$: $c=2$).}
\label{FIG-PhaseDiagram}
\end{figure}

In this section, the quantum phase diagram is summarized and discussed before the
underlying detailed 
analytical and numerical calculations are presented in the following sections.
The phase diagram in Fig.~\ref{FIG-PhaseDiagram} was derived by extensive density matrix renormalization group~(DMRG)\cite{White_1992,White_1993,Peschel_1999,Schollwoeck_2005} and infinite-size density matrix renormalization group (iDMRG) \cite{McCulloch_2008,Hu_2011,Hu_2014} methods with the $U(1)$ symmetry. OBC and shifted OBC (see appendix \ref{AppB})
are used unless stated explicitly otherwise.
The numerical calculations are implemented by keeping states in both blocks up to $2000$ for letting the truncation error less than $10^{-7}$. 
For the DMRG, larger than four iterative steps of sweep are used to guarantee the convergence of the ground state and the low-lying excited states. 
In the iDMRG simulation, a warm-up process with at least $1000$ truncated states is used and the number of states $m$ increases during the measurement.
The model Hamiltonian in Eq.~\eqref{HCHam} allows to define two parameters, 
which adjust the frustration $J_2/J_1$ in the legs 
and the coupling $J_{\perp}/J_1$ between the legs 
independently.  The resulting phase diagram in Fig.~\ref{FIG-PhaseDiagram} shows 
five extended phases: 
Haldane, NNN-Haldane, and staggered dimer (SD) for $J_{\perp}<0$, CD and RS
for $J_{\perp}>0$.  There is also a TLL phase along the line
of $J_{\perp}=0$ and $J_2<J_{2,c}$.

In the strong ferromagnetic limit of the odd-rung coupling $-J_{\perp}\gg J_1$, the 
spins on the odd rungs are in the triplet state, which effectively behave as spin-1.
The NN interaction $J_1$ to the spins in the rung-alternating model
may in principle 
cause interesting states with broken translational symmetry, where effective triplets over 
several rungs may be formed (so-called Haldane-dimer states).\cite{Amiri_2015,Hida_2014}
However, for the simple alternation of bonds in Eq.~(\ref{HCHam-int}) the states on the 
even rungs are always observed to be in an effective triplet state in agreement with
second order perturbation theory for $-J_{\perp}\gg J_1$.
In particular, for relatively strong NN coupling $J_1\gg J_2$ 
the triplets on the even rungs can form
valence bonds to the odd rungs on both sides equally, which 
corresponds to the well-known Haldane state.
For the spin-1 chain this 
state was shown to be topologically nontrivial and protected by time-reversal, parity and 
dihedral group symmetries (incomplete $Z_2\times Z_2$).\cite{Kjaell_2013,Pollmann_2012v2}
For OBC, topologically 
protected spin-1/2 degrees of freedom are localized at the edges.
Thus it is well characterized by a double degeneracy in the 
entanglement spectrum.\cite{Pollmann_2010}
It is also interesting that the translational symmetry by one-rung 
is recovered since no dimerization is observed in such a Haldane phase.
On the other hand, in the limit $J_2 \gg J_1$ a NNN-Haldane state is formed only involving
the odd rungs directly.   The spins on the even rungs then again form induced triplet states, 
which are also stable in a NNN-Haldane state.  In this case, the translation by one rung 
remains broken.  This state is therefore characterized by {\it two} weakly coupled Haldane 
states with valence bonds over NNN. 
The fractional edge spin-1/2
are no longer protected, so the state can adiabatically transit to a 
trivial state without breaking any symmetry.
Analogous to the $J_1$-$J_2$ spin-1 chain,\cite{Kolezhuk_1996,Kolezhuk_1997,Hikihara_2000,Kolezhuk_2002,Pixley_2014,Chepiga_2016_1,Chepiga_2016_2,Chepiga_2016_3}
a weak first order topological phase transition from the Haldane phase 
to NNN-Haldane phase can be observed.

For weaker ferromagnetic rung coupling there is a more complex competition between
frustration and triplet formation.  Above a critical value which can be predicted
from bosonization in the next section, the frustrating coupling $J_2$ causes a long-range
dimer order in each of the legs.  The weak ferromagnetic odd-rung coupling then causes
a staggered pattern of dimers on the two 
legs\cite{Metavitsiadis_2017} (SD phase).
In contrast to the Haldane phase this state is long-range ordered with a 
broken translational symmetry by one rung.  The region of the SD phase extends 
far to the ferromagnetic side, especially around $J_2\approx 0.6J_1$ 
where the dimerization and triplet gap of the underlying 
zigzag chains are largest.\cite{White_1996} 
However, the
 transition from Haldane to NNN-Haldane occurs at even larger $J_2$, which   
in turn causes an interesting reentrant
behavior around moderate values of $J_\perp \sim -1.5 J_1$:  Increasing $J_2$ causes 
a phase transition from Haldane to SD and then again from SD to Haldane states, before
finally the NNN-Haldane phase is observed at larger $J_2$.

In the limit of strong antiferromagnetic inter-leg coupling $J_{\perp}\gg J_1$, 
singlets are formed on the odd rungs.  Also on the even rungs an effective antiferromagnetic
correlation is induced via second order perturbation theory, so that a so-called 
rung singlet state\cite{Dagotto_1992,Barnes_1993,White_1994,Azzouz_1994,Dagotto_1996,Schmidt_2003} 
with short range correlations is found. Slightly stronger couplings along the 
legs cause resonating
valence bonds,\cite{White_1994} but do not change the nature of the correlations.
Therefore, in contrast to strong ferromagnetic rung couplings there are no degrees of 
freedom available which could allow another phase in the limit $J_{\perp}\gg J_1$.
For weak antiferromagnetic rungs, on the other hand, the dimerized state becomes important 
again.  For uncoupled legs dimerization is stable 
for $J_2>J_{2,c}$\cite{Affleck_1990,Okamoto_1992,Castilla_1995,Eggert_1996,White_1996} 
and the weak 
antiferromagnetic rungs now cause a columnar dimer 
pattern on the two legs.\cite{Metavitsiadis_2017}
The region of the resulting 
CD phase is again largest around $J_2\approx 0.6J_1$ where 
the strongest dimer correlations and triplet gaps are found.\cite{White_1996}
However, the CD phase only extends about a third in $|J_\perp|$
compared to the SD phase, which is simply due to the fact that breaking rung singlets 
costs three times the energy compared to the triplets.
For small $J_\perp$ the shape of the transition line can again be predicted by 
bosonization in the next section, which is found to be in 
the 2D Ising universality class.
Incommensurate short range correlations are found in the SD and CD phases for 
larger frustration $J_2$, but the dimerization order 
parameter changes continuously.\cite{White_1996}
The transition line between SD and CD phases has been 
drawn at $J_\perp=0$ in Fig.~\ref{FIG-PhaseDiagram}, since obviously the SD and CD pattern 
must be degenerate for uncoupled chains.  However, it must be emphasized that 
there is a large uncertainty if the CD phase may also be stable for small 
negative $J_\perp$
in a region around $J_2\sim 1.5 J_1$, which both bosonization and numerical results
seem to indicate 
as discussed in Secs.~\ref{SEC-BONSONIZATION} and \ref{SEC-NatTrans}.

\section{Field theory and renormalization group}\label{SEC-BONSONIZATION}
A RG treatment based on bosonization gives an analytical 
approach to determine the phase boundaries.\cite{Kim_2008,Starykh_2004,White_1996,Metavitsiadis_2014,Metavitsiadis_2017}
Here we employ non-Abelian bosonization to express the 
spin operators in an effective continuum theory of decoupled chains\cite{Affleck_1990, Eggert_1992,Ian1986409, tsvelik, book-cft}
\begin{equation}\label{spin operator}
\mathbf{S}(x)\approx\mathbf{J}(x)+(-1)^{x}~\Omega~\mathbf{n}(x)~,  
\end{equation} 
with $\Omega$  being a nonuniversal constant of order of one. 
The uniform part of the spin operator $\mathbf{J} =\mathbf{J}_L + \mathbf{J}_R$ is the sum of the chiral $\mathrm{SU}(2)$ currents of the Wess-Zumino-Witten (WZW) model,
while the staggered one is related to the matrix field $g$ of the WZW model via $\mathbf{n} \sim \mathrm{Tr} {\bm \sigma} g$, with ${\bm \sigma}$ the Pauli matrices.
By symmetry, the field  $\epsilon \sim \mathrm{Tr} g $  is also allowed corresponding to the dimerization operator,
$\epsilon_j = (-1)^j \mathbf{S}_j \cdot \mathbf{S}_{j+1}$.
%
In what follows we consider this theory for 
fixed points of decoupled chains, which are perturbed by the interchain couplings.
The RG 
treatment is then used to identify the leading instabilities, which in turn 
determine the phase transition lines.

Let us first consider the fixed point of $J_\perp=0$ and $J_2<J_{2,c}$, corresponding
to two decoupled gapless chains (black lines in Fig.~\ref{FIG_Model}).
It is well known that even without inter-leg coupling there is a marginal perturbation
from an in-chain current-current marginal operator\cite{Affleck1988} for each leg $\alpha=1,2$
\begin{equation}
\mathcal{H}_{a} =  2\pi v \int dx \lambda_{a} O_{a},~~O_{a} = \sum_{{\alpha}} \mathbf{J}_{
{\alpha},L} \cdot \mathbf{J}_{{\alpha},R}, \label{marginal}
\end{equation}
of scaling dimension $d_{a}=2$.
The velocity $v\approx {\pi J_1}/2 -1.65 J_2$ as well as the bare 
coupling $\lambda_{a} \approx 1.723(J_2 - J_{2,c})$ can be found as functions of the in-chain NNN coupling $J_2$.\cite{Affleck_1990,Eggert_1996,Eggert_1992}
The inter-leg couplings connect the chain field theories yielding a number of 
perturbations to the fixed point Hamiltonians, which in principle contain all 
operators allowed by symmetry.
The possible perturbing operators with the lowest scaling dimensions
are found to be 
\begin{equation} \label{hpert1}
\mathcal{H}_1 = 2 \pi v \int dx (\lambda_{\epsilon}O_\epsilon+\lambda_{n} O_{n}+\lambda_{c} O_{c} +\lambda_{b}O_{b})~,  
\end{equation}
with
\begin{eqnarray} \label{fp1-ops}
O_\epsilon &=& \epsilon_1 \epsilon_{2},~O_{n} = \mathbf{n}_1 \cdot \mathbf{n}_{2},~O_{c} = \mathbf{J}_1 \cdot \mathbf{n}_{2} + \mathbf{J}_2 \cdot \mathbf{n}_{1}\nonumber\\
O_{b} &=& \mathbf{J}_{1, L}\cdot \mathbf{J}_{2, R} + \mathbf{J}_{2, L}\cdot \mathbf{J}_{1, R}
 \end{eqnarray}
with scaling dimensions $d_\epsilon=d_n=1$,  $d_c=3/2$, and $d_b=2$.
The bare couplings are determined by using Eq.~\eqref{spin operator} 
%
\begin{equation}
\lambda_{\epsilon}=0,~~
\lambda_{n} = \Omega^2 \dfrac{J_\perp}{4\pi v},~~
\lambda_{c} = \Omega \dfrac{J_\perp}{4\pi v},~~
\lambda_{b} = \dfrac{J_\perp}{4\pi v}.
\end{equation} 
Note that although $\lambda_{\epsilon}$ is initially zero, it can be generated under the RG evolution of the couplings due to the second order contribution of the relevant $O_{c}$ operator.
\begin{figure}[t!]\label{figBosonization}
\includegraphics[width=0.95\columnwidth]{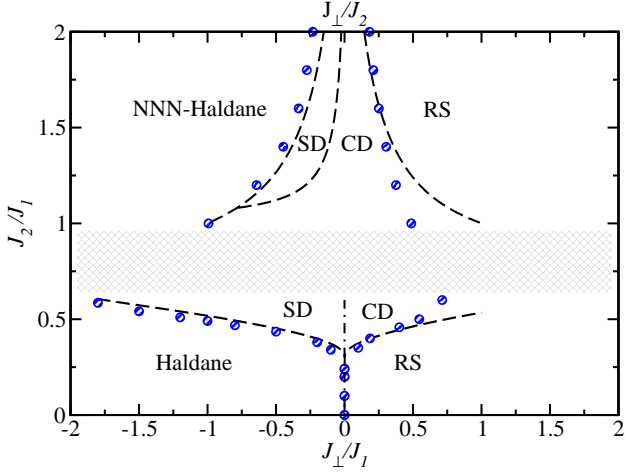}
\caption{
Ground state phase diagram of the honeycomb ladder obtained via the field theory and RG.  
The blue circles correspond to the critical points determined numerically via the 
DMRG algorithm, 
while the black dashed lines indicate the critical lines as determined via 
analytical RG using $\Omega = \pi $ and $\lambda_* = 1$.}
\label{figRGPhaseDiagram}
\end{figure}
In the RG treatment the energy cutoff 
is lowered according to $\Lambda(l)=\Lambda_0e^{-l}$ 
{with the dimensionless scale parameter $l$}, which changes the
effective coupling constants.\cite{Book_Cardy_1996, book-cft}
The evolution of each coupling is described 
by\cite{Book_Cardy_1996,tsvelik,Incollection_Senechal_2004}
\begin{equation}\label{couplingEvolution}
\frac{d\lambda_k}{dl} =(2-d_k)\lambda_k-\frac{\pi}{v}\sum_{i,j} C_{ijk} \lambda_i \lambda_j~, \quad
\end{equation}
where the coefficients $C_{ijk}$ are determined by the operator product expansion 
between the perturbing operators and $d_k$ is the corresponding scaling dimension.
In our case this results in the following RG equations (see appendix \ref{AppA})
\begin{subequations}
\label{rg-flow-a}
\begin{eqnarray}
\dot{\lambda_{a}}  &=&  \lambda_{a}^2 + \frac{1}{2}\lambda_{\epsilon}^2 - \frac{1}{2} \lambda_{n}^2, \label{rg-cpla} \\
\dot{\lambda_{b}}  &=&   \lambda_{b}^2 -\lambda_{\epsilon}\lambda_{n}
+ \lambda_{n}^2,
\label{rg-cplb} \\
\dot{\lambda_{\epsilon}} &=&  \lambda_{\epsilon} +\frac{3}{2}\lambda_{a}\lambda_{\epsilon}
-\frac{3}{2}\lambda_{b}\lambda_{n} +\frac{3}{2}\lambda_{c}^2, \label{rg-cple}\\
\dot{\lambda_{n}}  &=&  \lambda_{n} -\frac{1}{2}\lambda_{a}\lambda_{n}
-\frac{1}{2}\lambda_{b}\lambda_{\epsilon} +\lambda_{b}\lambda_{n} +\lambda_{c}^2, \label{rg-cpln}\\
\dot{\lambda_{c}}  &=&  \frac{1}{2}\lambda_{c} -\frac{1}{4}\lambda_{a}\lambda_{c}
+\lambda_{b}\lambda_{c} + \frac{1}{2}\lambda_{c}\lambda_{\epsilon} -\lambda_{c}\lambda_{n} \label{rg-cplc}.
\end{eqnarray}
\end{subequations}
We now integrate these equations until the magnitude of one coupling constant reaches
the cut-off $\lambda^*=1$, which in turn determines the dominant correlations.
As the strong coupling regime is reached, the system is no longer scale-invariant and 
renormalization stops.
When either $\lambda_{a}$ or $\lambda_{\epsilon}$ first reaches the strong coupling limit the ladder is dimerized, where 
$\lambda_{\epsilon}>0$ corresponds to an SD pattern while $\lambda_{\epsilon}<0$ gives the
CD phase.
When either $\lambda_{b}$ or $\lambda_{n}$ becomes dominant the system acquires 
strong correlations on the rungs, where $\lambda_{n}>0$ corresponds to singlets (RS) 
and  $\lambda_{n}<0$ to triplets (Haldane).
There is no region in the parameter space where the less relevant operator $O_{c}$ 
becomes dominant, and its role is restricted to simply promote one of the other 
phases.  If this operator can be made large by 
fine-tuning in an extended parameter space, 
it was argued to enhance incommensurate correlations.\cite{Metavitsiadis_2017}
The lower part of the phase
diagram in Fig.~\ref{figRGPhaseDiagram} was produced by identifying the coupling constant 
to first reach the values $\lambda_* = 1$ under integration of the RG equations, where
$\Omega = \pi$ was used for the best agreement with the DMRG results.
Both $\lambda_*$ and $\Omega$ are in principle adjustable parameters, which change the
exact location of the lines, but not the topology of the phases.

There is another possible fixed point in 
the limit $J_2 \gg J_1, J_\perp$, which consists of {\it four} decoupled chains 
denoted by $\bar \alpha = 1,2,3,4$
corresponding to the red lines in Fig.~\ref{FIG_Model}.
In this case, the couplings between the inner and the outer chains are different
depending on $J_1$ and $J_\perp$.
Consequently, we need to treat the marginal operator from Eq.~(\ref{marginal}) 
separately for the 
the inner chains ($\bar \alpha=2,3$), $O_{i}$ with coupling $\lambda_{i}$;
and for the outer ones ($\bar \alpha = 1,4$), $O_{o}$ with coupling $\lambda_{o}$.
The interchain perturbation now reads 
\begin{eqnarray} \label{hpert2}
\mathcal{H}_2 =  2 \pi v \int dx \Big[\lambda_{\epsilon}O_\epsilon&  +& \lambda_{n}O_{n} + \lambda_{b}O_{b}  + \lambda_{u}O_{u} \nonumber\\ 
&+&\lambda_{v} O_{v} +  \lambda_{w} O_{w} \Big], 
\end{eqnarray}
with,
\begin{eqnarray} \label{fp2-ops}
O_\epsilon &=& \epsilon_2 \epsilon_{3},~O_{n} = \mathbf{n}_2 \cdot \mathbf{n}_{3},\\
O_{b} &=& \mathbf{J}_{2, L}\cdot \mathbf{J}_{3, R} + \mathbf{J}_{3, L}\cdot \mathbf{J}_{2, R},\nonumber \\   
O_{u} &=&   \mathbf{J}_{1, L}\cdot \mathbf{J}_{2, R} + \mathbf{J}_{3, L}\cdot \mathbf{J}_{4, R} \ + \ L \leftrightarrow R , \nonumber\\ 
O_{v} &=&  \epsilon_{1} \partial_x{\epsilon}_{2}+ \epsilon_{4} \partial_x{\epsilon}_{3},~O_{w} = \mathbf{n}_{1} \cdot  \partial_x{\mathbf{n}}_{2}+ \mathbf{n}_{4} \cdot  \partial_x{\mathbf{n}}_{3},  \nonumber 
\end{eqnarray}
with scaling dimensions $d_\epsilon=d_n=1$ and $d_b=d_u=d_v=d_w=2$.
The bare couplings
\begin{eqnarray} \label{barecouplings-b}
\lambda_{\epsilon} &=& 0,~\lambda_{b} = \frac{J_\perp}{2 \pi v},~\lambda_{n} =~~~\Omega^2 \frac{J_\perp}{2 \pi v},\nonumber \\ 
\lambda_{v}&=&0,~\lambda_{u} = \frac{2J_1}{2 \pi v},~\lambda_{w} =-\Omega^2 \frac{J_1}{2\pi v }.  
\end{eqnarray}
where the velocity is $v\approx \pi J_2/2$.  The resulting RG equations are 
\begin{subequations} \label{rg-flow-b}
\begin{eqnarray}
\dot \lambda_{o} &=& \lambda_{o}^2 - \frac{1}{4}\left(\lambda_{v}^2-\lambda_{w}^2 \right),~\\
\dot \lambda_{i} &=& \lambda_{i}^2 + \frac{1}{2}\left(\lambda_{\epsilon}^2-\lambda_{n}^2 
\right)
-\frac{1}{4}\left(\lambda_{v}^2-\lambda_{w}^2 \right),~\\
\dot \lambda_{b} &=& \lambda_{b}^2 -\lambda_{n} \lambda_{\epsilon} +\lambda_{n}^2,~\\
\dot \lambda_{\epsilon} &=& \lambda_{\epsilon} + \frac{3}{2}\lambda_{i} \lambda_{\epsilon} - 
\frac{3}{2} \lambda_{b} \lambda_{n},~\\
\dot \lambda_{n} &=& \lambda_{n} -\frac{1}{2} \lambda_{i} \lambda_{n} -\frac{1}{2} \lambda_{b}
 \lambda_{\epsilon}  + \lambda_{b} \lambda_{n},~\\
\dot \lambda_{u} &=& \lambda_{u}^2 - \frac{1}{2} \lambda_{v} \lambda_{w} + \frac{1}{2} 
\lambda _{w}^2,~\\
\dot \lambda_{v} &=& \frac{3}{4}\left(\lambda_{o} + \lambda_{i}\right)\lambda_{v} -\frac{3}{2}
\lambda_{u} \lambda_{w},~\\
\dot \lambda_{w} &=& -\frac{1}{4} \left(\lambda_{o} + \lambda_{i}\right) \lambda_{w} -
\frac{1}{2} \lambda_{u} \lambda_{v} + \lambda_{u} \lambda_{w}.~
\end{eqnarray}
\end{subequations}
For weak perturbations we find that the critical lines are determined by 
the competition of the 
relevant $O_{n}$,
driving the system into strong rung correlations, and 
the marginal $O_{u}$, $O_{w}$ both of which drive the system into a dimerized phase for large 
values.  The dimer operator $O_{\epsilon}$ is more relevant than the marginal couplings
but remains smaller under RG flow for the starting values in Eq.~(\ref{barecouplings-b}).
However, the pattern (SD or CD) is still determined by the sign of $\lambda_\epsilon$
which indicates that the transition line is shifted to {\it negative} values of $J_\perp$ 
in Fig.~\ref{figRGPhaseDiagram}.
It should be noted that the numerical analysis of the exact location of the
transition line is also extremely
difficult which will be discussed below in Sec.~\ref{SEC-NatTrans}, so it
remains an open question if the CD phase may be stable for small negative $J_\perp$.
All other transition lines 
are in
quantitatively good agreement in the vicinity of both fixed points in Fig.~\ref{figRGPhaseDiagram}.  For larger bare couplings the RG approach breaks down, so no meaningful prediction can 
be made with bosonization 
in the shaded region of Fig.~\ref{figRGPhaseDiagram}, which separates the analysis 
at the two fixed points.
\section{Nature of the Transitions}\label{SEC-NatTrans}
This section is devoted to discuss detailed numerical results for
the nature of phase transitions presented in Sec.~\ref{SEC-PhaseDiagram}.
Traditional phase transition can be fully described in the framework of Landau theory of symmetry breaking, where the concept of \textit{local} order parameter is involved.
The critical exponents for the continuous phase transition can be extracted from the order parameter.
However, a so-called symmetry protected topological (SPT) state, 
may only be characterized by \textit{non-local} order parameter dubbed as topological order.\cite{Pollmann_2012}
Since all the other disordered phases are adjacent to the dimerized phases in the parameter space considered,
one can use the local order parameter of the CD/SD phase to distinguish the 
phase boundaries and extract the critical exponents.
The dimerization order parameters in the columnar and staggered patterns 
$D_L^{\textrm{c}}$ and $D_L^{\textrm{s}}$, respectively are given by
\begin{equation}\label{OPCD}
D^{\rm c/s}_{L}(j) = \sum_{\alpha=1,2}\left\vert\frac{(\pm 1)^{\alpha}}{2}
\left[\langle \textrm{S}_{j,\alpha}\cdot 
\textrm{S}_{j+1,\alpha}\rangle - \langle \textrm{S}_{j-1,\alpha}\cdot 
\textrm{S}_{j,\alpha}\rangle\right]\right\vert.
\end{equation}
The dimerization shows strong 
{\it Friedel} oscillations near edges, which
 decay exponentially towards the middle.
Therefore we shall set $j=L/2$ in Eq.~\eqref{OPCD} so as to minimize the boundary effect and
chose compatible OBC as discussed in appendix \ref{AppB}.
%

%
%
\subsection{RS--CD transition}
\begin{figure}[t]
\centering
\includegraphics[width=0.95\columnwidth, clip]{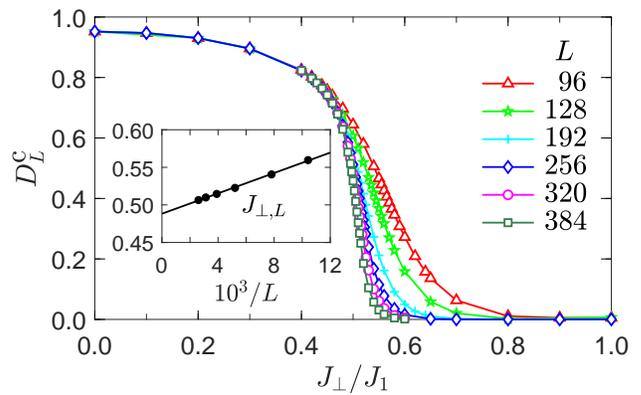}\\
\caption{Columnar dimerization order parameter $D_L^{\textrm{c}}$ for OBC as a function of $J_{\perp}$ for $L=96$~($\Delta$), $128$~(\FiveStarOpen), $192$~($+$), $256$~($\diamondsuit$), $320$~(\Circle) and $384$~($\Box$)
when $J_2=J_1$ is fixed. In the inset, we do a linear extrapolation of the quasi-critical points $J_{\perp,L}$ ($\bullet$) to the TDL.}\label{FIG-OPDCgt0}
\end{figure}

The typical behavior of the order parameter $D_L^{\textrm{c}}$ with 
increasing $J_{\perp}$ at fixed $J_2=1.0$ is presented in Fig.~\ref{FIG-OPDCgt0} for
different lengths $L$ and OBC.
%
To determine the accurate value of the critical point $J_{\perp,c}$, we shall first obtain quasi-critical points $J_{\perp,L}$ which are defined by the position of 
the first-order derivative peaks of $D_L^{\textrm{c}}$ for each length $L$.
It is then possible to make an extrapolation to the thermodynamic limit (TDL) to find
$J_{\perp,c}$ as shown in the inset of Fig.~\ref{FIG-OPDCgt0}, which gives
$J_{\perp,c}\simeq0.4883(12)$ in this case.
The signature of the phase transition can also be 
seen by local entanglement measures\cite{Campos_Venuti_2006} and by the 
rung-singlet correlation in the RS phase.\cite{Wang_2000}

\begin{figure}[t]
\centering
\includegraphics[width=0.95\columnwidth, clip]{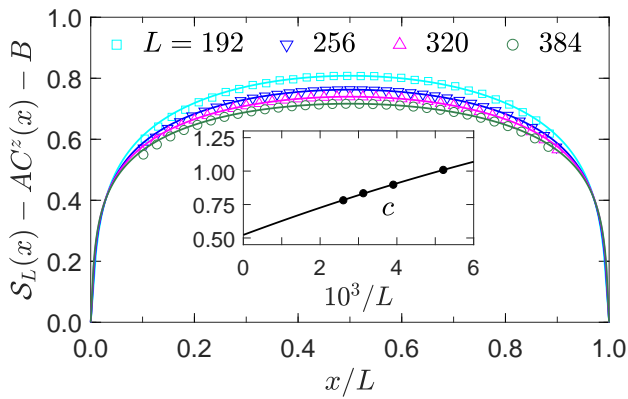}\\
\caption{Von Neumann entropy $\mathcal{S}$ as a function of partition size $x/L$ fitted by Eq.~\eqref{vNEOBC} at CD-RS quasi-critical points $J_{\perp,L}$ for different $L$.
The inset shows a second-order polynomial extrapolation of the central charge 
to $c=0.52(3)$ in the TDL.}\label{FIG-RSvsCD-CC}
\end{figure}

Fundamental characteristics of ground states in quantum many-body systems are 
manifest in their entanglement properties. In particular, 
the von Neumann entropy has drawn much attention in condensed matter systems,
which exhibits a universal scaling behavior and 
captures some criticality features.\cite{Osterloh_2002,Wu_2004,Tagliacozzo_2008,Latorre_2009,Laflorencie_2016}
For example, the \textit{central charge} describes the universality class of the criticality and sub-leading corrections, for critical points in one-dimensional quantum models.\cite{Tagliacozzo_2008,Latorre_2009,Laflorencie_2016}
For the frustrated honeycomb ladder 
we use the scaling of the von Neumann entropy $\mathcal{S}_L(x)$ 
\begin{equation}\label{vNEOBC}
\mathcal{S}_L(x) = \frac{c}{6}\ln\left[\frac{2L}{\pi}\sin\left(\frac{\pi x}{L}\right)\right] + AC^{z}(x) + B,~
\end{equation}
where $c$ is the central charge, $A$ and $B$ are non-universal fitting parameters, and
$x$ is the size of the partition from the edge in units of the lattice spacing. 
Here $C^{z}(x) = C_{\alpha}^{z}(x) \equiv \langle S_{x,\alpha}^zS_{x+1,\alpha}^z \rangle$ is the $z$-component of the spin-spin correlation function along the leg direction, which is
independent of the leg index $\alpha=1,\;2$
and alternates due to OBC.
Its relation to the alternating term of the von Neumann entropy under OBC was firstly noted by Wang,\cite{Wang_2004} and then demonstrated numerically by Laflorencie \textit{et al.}\cite{Laflorencie_2006}
Recently, the fluctuations of the spin-spin correlation function for the spin-1/2 $XX$ chain were calculated analytically, which show that the sub-leading terms of the oscillations
exist.\cite{Song_2010,Song_2012}
In order to obtain a rather accurate central charge $c$ defined in Eq.~\eqref{vNEOBC}, some tricks can be employed to reduce the finite size effects and the rapid oscillations due to the OBC as well as the alternating nature of the honeycomb ladder.
We calculate the central charge at the quasi-critical points $J_{\perp,L}$ for each length $L$ and extrapolate, as shown in Fig.~\ref{FIG-RSvsCD-CC}.
We stress here that each quasi-critical point $J_{\perp,L}$ happens to be the position of the maximum value of von Neumann entropy.
The resulting central charge $c\simeq0.52(3)$ indicates
that the phase transition RS--CD has the universality class of the 
2D Ising type with an exact central charge of $c=1/2$.\cite{book-cft}           

\begin{figure}[t]
\centering
\includegraphics[width=0.95\columnwidth, clip]{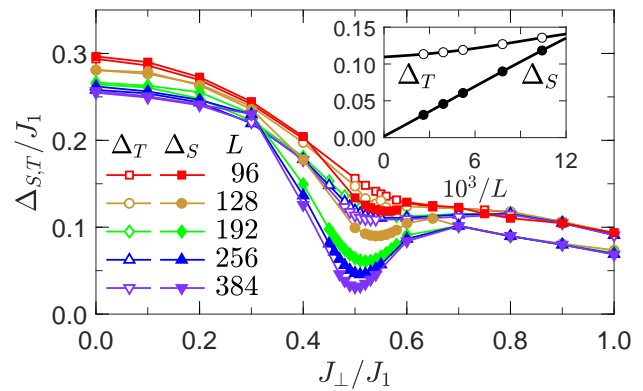}\\
\caption{Evolution of singlet gap $\Delta_{S}$~(filled symbols) and triplet gap $\Delta_{T}$~(open symbols) as a function of $J_{\perp}$ when $J_2=1.0$ for $L=96$, $128$, $192$, $256$ and $384$. The inset gives a linear fit of $\Delta_{S}$ ($\bullet$) and a second-order polynomial fit $\Delta_{T}$ ($\circ$) at the quasi-critical points to the TDL.}\label{FIG-SpinGapJ2100P}
\end{figure}

We now take a closer look at the energy gaps near the phase transitions.  
The ground state can always be found in the sector with total spin-$z$ $S^{z}_{t}=0$.
The triplet gap 
$\Delta_T$ 
is determined numerically by the
energy difference to the first excited state with $S^{z}_{t}=1$.
We can also find the singlet gap $\Delta_S$
to the first excited state with $S^{z}_{t}=0$. 
We show the value of gaps for various sizes $L$ in Fig.~\ref{FIG-SpinGapJ2100P}, where 
for each $L$ the local minimum of the singlet 
gap corresponds to the quasi-critical point $J_{\perp,L}$ in Fig.~\ref{FIG-RSvsCD-CC}. 
An extrapolation of the gaps at quasi-critical points to the TDL is shown 
in the inset.
When $J_{\perp}$ crosses the critical point, the triplet gap remains finite while the singlet gap tends to be zero.
\subsection{Haldane/NNN-Haldane--SD transition}
\begin{figure}[t]
\centering
\includegraphics[width=0.95\columnwidth, clip]{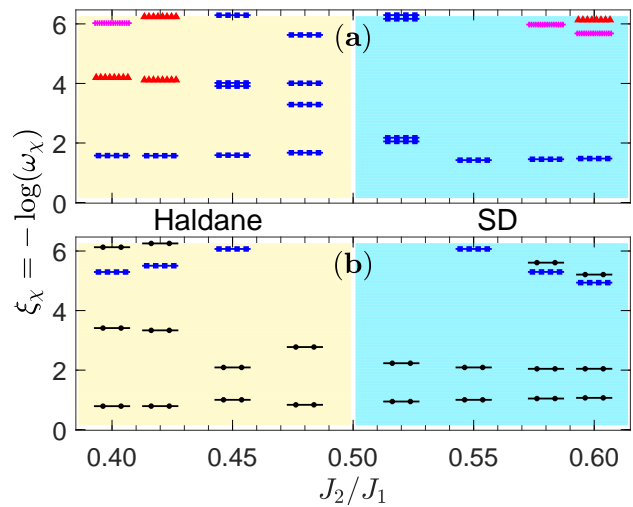}\\
\caption{Lower part of the entanglement spectrum of the Haldane and SD phases in the vicinity of the critical point $J_{2,c}\simeq0.5$ at fixed $J_{\perp}=-1.0$ for 
(a) $L=64$ with PBC (b) $L=95$ with shifted OBC~(see appendix \ref{AppB}). 
 In the multiplet structure the degeneracy is indicated by different symbols:
 $2$~(black circles), $4$~(blue squares), $8$~(red triangles) and $16$~(magenta plus).
}\label{FIG-ESJvM100}
\end{figure}

The transition between the Haldane/NNN-Haldane phase and the SD phase is 
similar to that of the RS--CD transition.
The phase transition could be found as before, but instead we want to take the opportunity
to illustrate that a finite-size-scaling~(FSS) method with critical indices gives
a good data collapse.
\begin{figure}[t]
\centering
\includegraphics[width=0.95\columnwidth, clip]{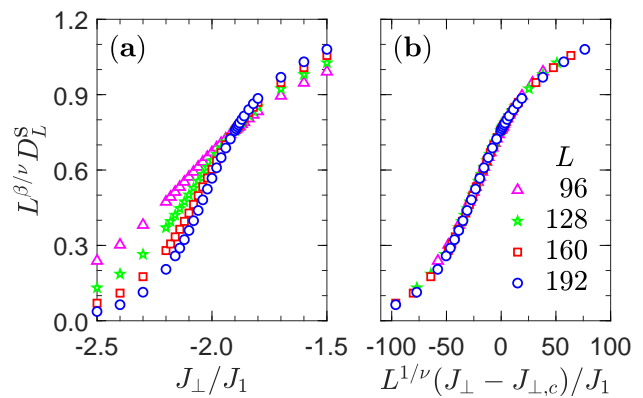}\\
\caption{FSS analysis of the staggered dimerization order parameter $D_L^{\textrm{s}}$ 
for $J_2=0.7J_1$ as a function of $J_{\perp}$ for different length $L=96$~($\Delta$), $128$~(\FiveStarOpen), $160$~($\Box$) and $192$~(\Circle). The best fitting suggests $J_{\perp,c}=-1.8980(5)$ and critical exponents $\beta=0.126(2)$ and $\nu=0.98(2)$.}\label{FIG-NNNHvsCD-FSS}
\end{figure}
In particular, the behavior of the order parameter $D^{s}_{L}$ with the finite size $L$ 
follows
\begin{equation}\label{FSS}
D^{s}_{L}(J) \simeq L^{-\beta/\nu}f_D\left(\vert J-J_c\vert L^{1/\nu}\right),
\end{equation}
where the critical exponent $\nu$ of the gap describes the divergence of the correlation length while $\beta$ is the critical exponent of the order parameter $D\sim\vert J-J_c\vert^{\beta}$.
To obtain critical exponents, we change exponents $\mu_{1,2}$ until we see 
the collapse of $D^{s}_{L}(J)L^{\mu_{1}}$ as a function of $\vert J_{\perp}-J_{\perp,c}\vert^{\mu_2}$ for all lengths $L$.
In this manner we extract $\beta = \mu_1/\mu_2$ and $\nu = 1/\mu_2$.
As an example, the FSS of the order parameter $D_L^{\textrm{s}}$
for $J_{2}=0.7$ is shown in Fig.~\ref{FIG-NNNHvsCD-FSS} for the SD-Haldane transition.
From the scaling of $D_L^{\textrm{s}}$ we obtain the critical point $J_{\perp,c}/J_1\simeq-1.8980(5)$, critical exponents $\beta=0.126(2)$ and $\nu=0.98(2)$.
These values coincide well with the critical indices of the 
2D Ising universality class where $\beta=1/8$ and $\nu=1$.
Following the same way introduced in the previous subsection, we get the central charge at the critical point $J_{\perp,c}$, $c=0.51(2)$, which is another independent evidence of the 2D Ising universality class.
\begin{figure}[t]
\centering
\includegraphics[width=0.95\columnwidth, clip]{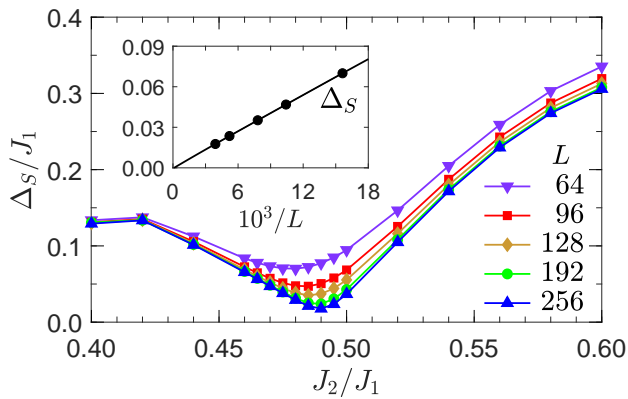}\\
\caption{Evolution of the singlet gap for shifted OBC 
in the vicinity of the critical point $J_{2,c}\simeq0.5$ at fixed $J_{\perp}=-1.0$ for $L=64$~($\nabla$), $96$~($\Box$), $128$~($\diamondsuit$), $192$~(\Circle) and $256$~($\Delta$). In the inset, a linear extrapolation of the singlet gap $\Delta_{S}$ at quasi-critical points goes to zero in the TDL.}\label{FIG-SpinGapJvM100}
\end{figure}
We now turn to the topological properties of the Haldane and SD phases by analyzing the
entanglement spectrum $\xi_{\chi}$,
which is derived from DMRG eigenvalues $w_{\chi}$ of the reduced density-matrix 
$\rho_{A}$ for the right half of the system 
if the degrees of freedom in the other half are traced out, 
namely $\xi_{\chi}=-\ln(w_{\chi})$.\cite{Li_2008}
It has been shown that the entanglement spectrum of a SPT state exhibits a nontrivial multiplet structure of even-fold degeneracy.\cite{Pollmann_2010}
However, the multiplet structure is influenced by the ladder structure and is 
strongly dependent
on the boundary condition.
For example, the one-dimensional spin-$1$ Heisenberg chain is gapped with a 
unique ground state and a four-fold degeneracy of the entanglement spectrum under PBC.
For OBC the edge states give rise to a four-fold degenerate ground state in the TDL and
a two-fold degeneracy of the entanglement spectrum.\cite{Pollmann_2010}
For the honeycomb ladder the choice of OBC is therefore quite important as also
discussed in appendix \ref{AppB}.
%
In both SD and Haldane phases with OBC, the ground state would generate 
two spin-1/2 confined at the edges.  By a relative shift of the legs 
by one site (shifted OBC) it is possible to provide additional edge spins as discussed in 
appendix \ref{AppB}.  In this case the ground state is unique and the 
entanglement spectrum has no spurious degeneracy.
This can be verified by the nonzero spin singlet gap $\Delta_S$ for an 
interchain coupling $J_{\perp}=-1.0$ as a function of NNN coupling $J_2$ in 
Fig.~\ref{FIG-SpinGapJvM100}.
Near the critical point an \textit{avoided level crossing} phenomenon emerges, where 
the gap shows a local minimum with linear finite size behavior towards zero as shown in the inset. 
The entanglement 
spectrum recovers an even-fold degeneracy as shown Fig.~\ref{FIG-ESJvM100}(b).
In Fig.~\ref{FIG-ESJvM100}(a) the entanglement spectrum for PBC is shown.
Remarkably, the degeneracy of the entanglement spectrum in the SD phase is also 
even-fold and therefore we cannot distinguish the two phases 
solely through the entanglement spectrum.
\subsection{Haldane--RS transition}
\begin{figure}[t]
\centering
\includegraphics[width=0.95\columnwidth, clip]{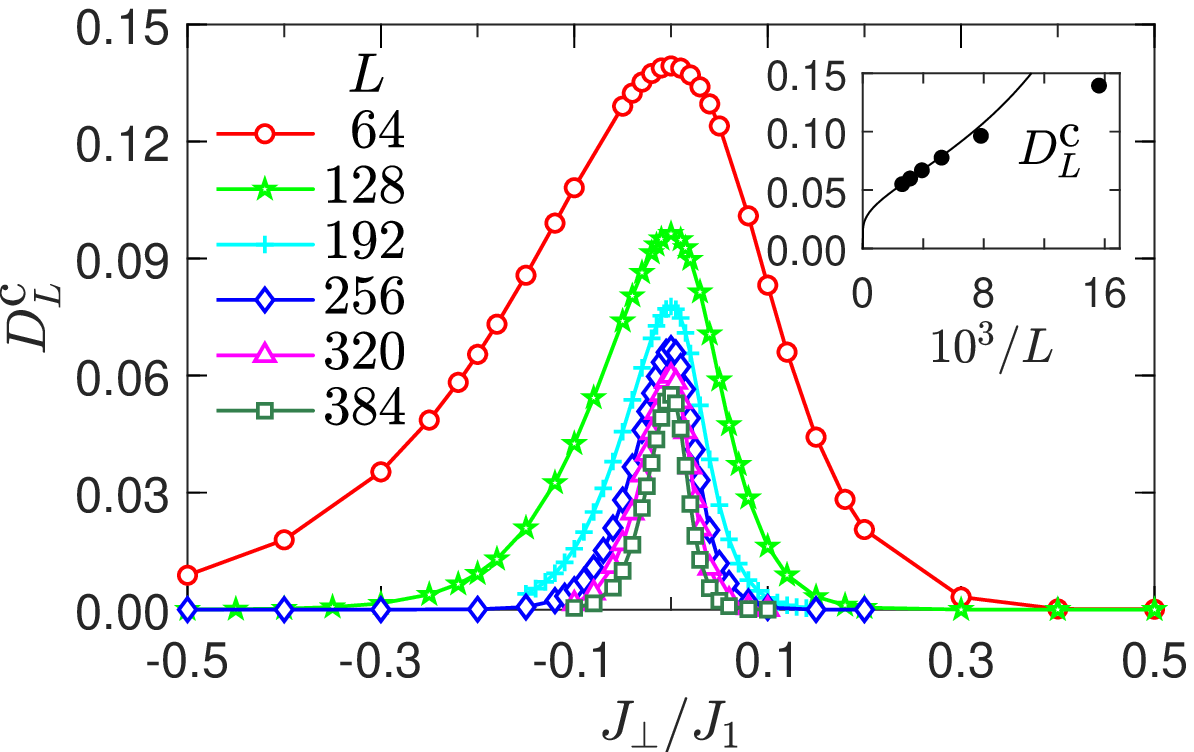}\\
\caption{Columnar dimerization order parameter $D_L^{\textrm{c}}$ as a function of $J_{\perp}$ at fixed $J_2=0.2$ for $L=64$~(\Circle), $128$~(\FiveStarOpen), $192$~($+$), $256$~($\diamondsuit$), $320$~($\Delta$) and $384$~($\Box$). Inset shows the extrapolation of maximal values 
using $D_L^{\textrm{c}}\to 0.1268/\ln(L/38.5)$ as $L\to \infty$.}\label{FIG-CDOP-J2020}
\end{figure}

Haldane and RS phases are both disordered and uniform without breaking any 
time-reversal, parity and point group symmetries.
Here the Haldane phase is a topologically nontrivial state with finite 
string order and edge spins,
while the RS state does not have edge spins, which is due to the 
different symmetry of the corresponding string order parameter.\cite{Kim_2008}
The direct transition between the Haldane phase and RS phase is continuous with central 
charge $c=2$, which has been the topic of many works.\cite{Dagotto_1992,Barnes_1993,White_1994,Azzouz_1994,Dagotto_1996,Schmidt_2003,Starykh_2004}
To show that there is not an intermediate dimerized phase for $J_2<J_{2,c}$, we illustrate the behavior of the order parameter of the dimerized phase across the transition in Fig.~\ref{FIG-CDOP-J2020} for $J_2=0.2J_1$.
Without loss of generality, hereafter we shall only consider OBC 
and let $J_2= 0.2J_1$ throughout the subsection.
For each given length $L$, the maximal value ${D_L^{\textrm{c}}}$ is always
exactly at $J_{\perp}=0$.
The extrapolation of quasi-critical points $J_{\perp,L}$ on both sides
show that the critical point is $J_{\perp}=0$ within the limit of error.
The maximal values ${D_L^{\textrm{c}}}$ in the inset approach the TDL on a 
non-linear curve, which is consistent with a logarithmically slow
$D^c_L\sim 1/\ln(L/L_0)$ decrease as $L\to \infty$
in agreement with the RG flow of the corresponding
marginal operator in Eq.~(\ref{marginal}).\cite{Affleck_1990,Okamoto_1992,Castilla_1995,Eggert_1996,White_1996,Eggert_1992,Sorensen_1993}
%
%

\begin{figure}[t]
\centering
\includegraphics[width=0.95\columnwidth, clip]{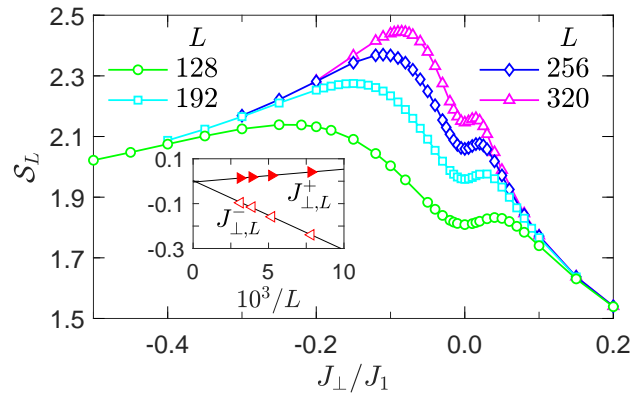}\\
\caption{Von Neumann entropy $\mathcal{S}_{L}$ as a function of $J_{\perp}$ at fixed $J_2=0.2$ for $L=128$~(\Circle), $192$~($\Box$), $256$~($\diamondsuit$) and $320$~($\Delta$). In the inset, linear extrapolations of peak-positions $J^{\pm}_{\perp,L}$ on both sides approach 
the same critical point $J_{\perp,c}=0$ in the TDL.}\label{FIG-VNE-J2020}
\end{figure}

%
The phase transition is also reflected in the von Neumann entropy $\mathcal{S}_L$, which is shown in Fig.~\ref{FIG-VNE-J2020} for different lengths $L$ of the system.
For each given $L$, $\mathcal{S}_L$ has a peak for the antiferromagnetic~($J_{\perp}>0$) and ferromagnetic~($J_{\perp}<0$) coupling, respectively.
At the decoupled case~($J_{\perp}=0$), a local minimum emerges.
With increasing length $L$, positions of the peaks $J^{\pm}_{\perp,L}$ on both sides converge to the symmetric 
point $J_{\perp,c}=0$ where the continuous phase transition occurs (see inset).
We also obtain the central charge $c$ from the von Neumann entropy $\mathcal{S}_L$.
Best fitting with Eq.~\eqref{vNEOBC} for different sizes $L$ at the point $J_{\perp}=0$ suggests $c=1.98(3)$, which is in agreement with the existence of
two decoupled TLL with $c=1$ each for $J_\perp=0$.

\subsection{Haldane--NNN-Haldane transition}
The spin-1 chain with NNN interactions was studied in previous works,\cite{Kolezhuk_1996,Kolezhuk_1997,Hikihara_2000,Kolezhuk_2002,Pixley_2014,Chepiga_2016_1,Chepiga_2016_2,Chepiga_2016_3} where a weak first order transition
between the topological Haldane phase and the NNN-Haldane phase was found.
The situation in the honeycomb ladder with the strong-ferromagnetic odd rungs 
is slightly different, however, since triplet correlations on the even rungs are only
indirectly induced by 
second order perturbation theory.  Therefore, a careful analysis of this phase transition
must be made to exclude the possibility of intermediate phases.\cite{Amiri_2015,Hida_2014}

\begin{figure}[t]
\centering
\includegraphics[width=0.95\columnwidth, clip]{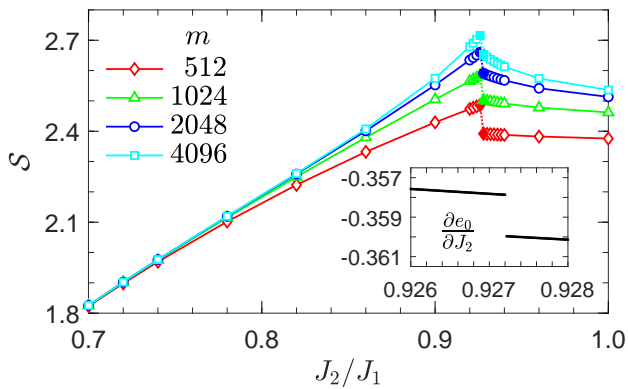}\\
\caption{Von Neumann entropy as a function of $J_{2}$ when fixed $J_{\perp}=-3.0$ for the Haldane-NNN--Haldane phase transition. We use $m=512$~($\diamondsuit$), $1024$~($\Delta$), $2048$~(\Circle) and $4096$~($\Box$) in the iDMRG calculations. The inset shows the first-derivative of the average energy $e_{0}$ per site with respect to $J_{2}$ when $m=4096$.
}\label{FIG-HDvsNNNHD-VNE}
\end{figure}

We now use the iDMRG method,\cite{McCulloch_2008,Hu_2011,Hu_2014} where no 
finite-size effect is involved.
Instead of finite size scaling, an extrapolation in the number of kept state $m$ is
useful so we vary this control parameter from $512$ to $4096$.
The behavior 
of the von Neumann entropy $\mathcal{S}$ with frustration $J_2$ for 
different kept states $m$ is shown in Fig.~\ref{FIG-HDvsNNNHD-VNE}.
The entropy has a discontinuous behavior in the narrow regime $J_2\in(0.926, 0.928)$ 
in all cases, which clearly signals a phase transition.
To further pinpoint the transition point, we analyze the derivative 
of the energy from the left and from the right side in this region as shown in
the inset, which shows a clear jump at
 the transition point $J_{2,c}\simeq0.92720754$.
We therefore find no sign of intermediate phases or continuous behavior, so  that the
transition is of first order.
\subsection{SD--CD phase transition} \label{SDCD}
\begin{figure}[t]
\centering
\includegraphics[width=0.95\columnwidth, clip]{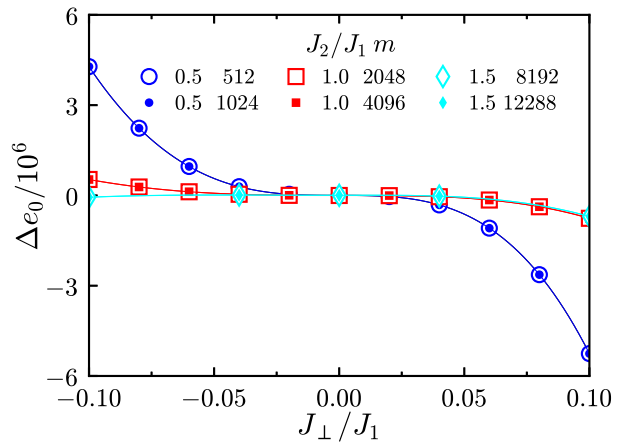}\\
\caption{Energy difference $\Delta e_{0}$ per site between OBC and shifted OBC
given by iDMRG calculations. We choose three typical cases of $J_2=0.5$~(\Circle), $1.0$~($\Box$) and $1.5$~($\diamondsuit$) respectively. For each case, different truncated dimension in iDMRG is used from $m=512$ up to $12,288$. The error bar is smaller than the symbol size.}\label{FIG-SD-CD}
\end{figure}

SD and CD phases are both dimerized states with conventional order parameters 
in the framework of Landau theory of phase transitions.
Since they break different point group symmetries, the transition between them should 
be first-order.
However it is a surprisingly big challenge to determine the transition line 
just from the numerical calculations.
The main reason for this is that the observed dimerization pattern in finite chains
also strongly depends on the 
edge geometries of the OBC or shifted OBC (see appendix \ref{AppB}).
To minimize the edge effect we therefore use
the iDMRG method again.  During the warm-up process both OBC and shifted OBC are used,
until the average ground-state energy per site $e_{0}$ reaches 
very good convergence.
Very close to $J_{\perp}=0$ the shifted OBC always result in an SD phase, while OBC
give the CD phase, albeit with slightly different energies $e^{s}_{0}$ and $e^{c}_{0}$,
 respectively.
To identify the true ground state at zero temperature, we then 
analyze the energy difference $\Delta e_{0}=e^{c}_{0}-e^{s}_{0}$ per site.
In Fig.~\ref{FIG-SD-CD}, we show the results for three 
typical cases of $J_{2}=0.5$, $1.0$ and $1.5$ and different states kept $m$.
When $J_{2}=0.5$ and $1.0$, the convergent $\Delta e_{0}$ always gives a positive sign on the left side ($J_{\perp}<0$) and a negative sign on the right side ($J_{\perp}>0$), so that
the SD--CD transition appears to occur exactly at $J_{\perp}=0$.
At $J_{2}=1.5$, the energy difference becomes very small and more
than $10,000$ states are required for convergence.  The energy difference 
remains negative even for $J_\perp/J_1 = -0.1$, but a clear 
position for the phase transition cannot be determined.
It is also known that the dimerization becomes exponentially small for 
large $J_2$.\cite{White_1996}
At this point it is unclear, if the phase transition remains at $J_\perp=0$ or if 
there is a region where the CD phase is stable at
negative values of $J_\perp$ as also suggested by the 
bosonization analysis in Sec.~\ref{SEC-BONSONIZATION}, 
which will be an interesting but challenging topic for future research.

\section{Energy Gaps}\label{SEC-SpinGap}
\subsection{A Special Case: $J_{2}=0$}
\begin{figure}[t]
\centering
\includegraphics[width=0.95\columnwidth, clip]{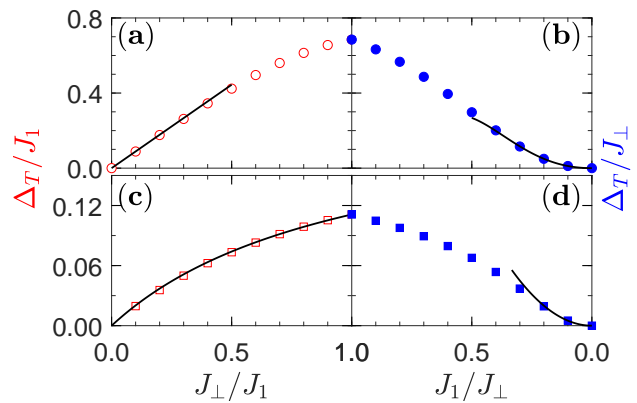}\\
\caption{Triplet gap $\Delta_{T}$ of the honeycomb cell (a,b) and the honeycomb ladder (c,d)
for $J_2=0$ compared to strong coupling fits (solid lines).}
\label{FIG-SpinGapSCELLvsTDL}
\end{figure}

The honeycomb ladder is unique in 
that the couplings on even rungs are absent.
Therefore, the correlations on the even rungs are only indirectly induced in second order
 corresponding to a weak energy gain.  This has interesting consequences for the energy gaps.
For example, in the ordinary ladder it is known that the triplet gap $\Delta_{T}$ increases
linearly with larger rung coupling,\cite{Reigrotzki_1994,Cabra_1997}
while we find that it 
decreases for stronger $J_{\perp}$ in the honeycomb ladder as shown in 
Fig.~\ref{FIG-SpinGapSCELLvsTDL}(d) for $J_2=0$.
This behavior can in fact be understood by considering the subsystem of 
three adjacent rungs, which simply forms a ring of 6 spins coupled by $J_1$ and $J_\perp$
 (honeycomb cell).  The behavior of the triplet gap is analogous to the honeycomb ladder
as shown in Fig.~\ref{FIG-SpinGapSCELLvsTDL}.

For $J_1 \gg J_\perp$ doublet states are formed by the top three and bottom three
spins in the honeycomb cell, which in turn are coupled by $J_\perp$, so the 
triplet gap increases linearly with $\Delta_{T}\simeq{8} J_{\perp}/9$ 
in Fig.~\ref{FIG-SpinGapSCELLvsTDL}(a).  For the honeycomb ladder we also find a linear 
behavior, which fits well to the expression 
\begin{equation}\label{HCLadderSpinGapWCL}
\frac{\Delta_{T}}{J_{1}} \approx 2\Delta_0\frac{J_{\perp}}{J_{\perp}+J_1}
\end{equation}
with $\Delta_0=0.1110382(1)$.

In the opposite limit of $J_1\ll J_{\perp}$ the singlet-triplet gap in the honeycomb cell 
is produced by an indirect coupling of the two spins of the central (uncoupled) rung, 
which can be found by a perturbation expansion
\begin{equation}\label{HCCellSpinGapSCL}
\frac{\Delta_{T}}{J_{\perp}} = \lambda^2\left[1+\frac{3}{2}\lambda-\frac{3}{8}\lambda^2-\frac{75}{16}\lambda^3+\mathcal{O}(\lambda^4)\right]
\end{equation}
where $\lambda=J_{1}/J_{\perp}$.  This also explains the drop of
the triplet gap in the honeycomb ladder for large $J_\perp$, which follows
$\Delta_{T} \simeq \frac12{J_1^2}/{J_{\perp}}$ in that limit.

\subsection{$J_2>0$}
\begin{figure}[t]
\centering
\includegraphics[width=0.95\columnwidth, clip]{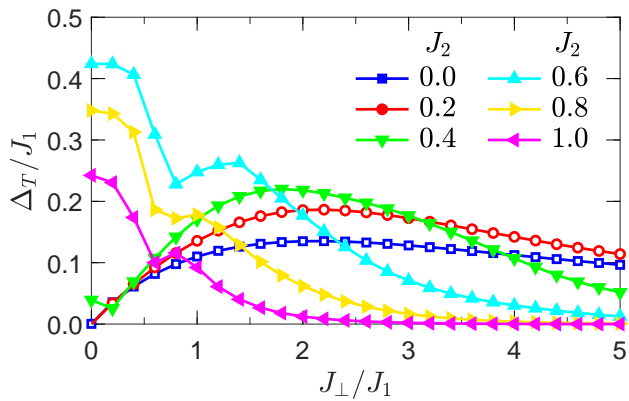}\\
\caption{Evolution of the triplet gap $\Delta_{T}$ for different $J_2$
in the TDL.  }\label{FIG-SpinGapJ20TO1}
\end{figure}

We now turn on the frustrating coupling $J_2$ to see if the non-monotonic behavior 
of the triplet gap with $J_\perp$ changes.
The numerical results of different $J_2$ are shown in Fig.~\ref{FIG-SpinGapJ20TO1}.
For small $J_2<J_{2,c}$, the 
quick increase and slow drop-off with $J_\perp$ gives a broad maximum 
analogous to the $J_2=0$ case above.
On the other hand, for larger $J_2> J_{2,c}$ the triplet gap is already nonzero 
for $J_\perp=0$ and then first {\it drops} to a minimum before rising and falling again.
The size and location of the minimum is directly related to value of the gap at 
$J_\perp=0$ which is known\cite{White_1996} to have a maximum around $J_2=0.6J_1$.
In fact, the positions of the gap minima roughly coincide with the CD-RS transition line,
which implies that $J_\perp$ reduces the gap in the CD phase, while the broad maximum
is a signature of the RS phase.
Moreover, in the RS phase at a given $J_\perp$ a maximum can always be 
observed as a function of $J_2$.

\section{Conclusions}\label{SEC-Con}
In summary, we systematically analyzed the spin-1/2 honeycomb ladder as a function of
inter-leg coupling $J_\perp$ on the odd rungs and 
NNN frustration on the legs $J_2$.  Detailed numerical results for the order parameters,
the entanglement spectrum, and the energy gaps from DMRG and iDMRG give a complete
picture of the phase transitions, which agree quantitatively with our analytical 
bosonization calculations.

For very strong ferromagnetic odd-rung couplings we find a weak first order transition as a 
function of $J_2$ between
Haldane and NNN-Haldane phases analogous to the frustrated spin-1 chain.\cite{Kolezhuk_1996,Kolezhuk_1997,Hikihara_2000,Kolezhuk_2002,Pixley_2014,Chepiga_2016_1,Chepiga_2016_2,Chepiga_2016_3}
However, the SD phase with staggered dimer order is stable for relatively large 
ferromagnetic odd rungs which pushes far into the Haldane phase
and leads to a reentrant behavior as a function of $J_2$.
For antiferromagnetic odd-rung coupling we find a transition from the CD phase with
columnar dimer order
to a RS phase with rung singlets 
as a function of $J_\perp$ and again a reentrant behavior as a function
of $J_2$ which can be linked to the corresponding  
maximum of the triplet gap in the underlying zigzag chain around $J_2\approx 0.6J_1$.\cite{White_1996}  The transitions from the ordered dimer phases are always of the 2D Ising
universality class.

The line $J_\perp=0$ is critical below $J_2<J_{2,c}$ and corresponds to a $c=2$ transition
between Haldane and RS phases.  For larger $J_2$ the CD and SD phases appear, which must be
degenerate for $J_\perp=0$.  However,
 numerical and bosonization results surprisingly find that 
the CD phase may be stable also for small negative values of $J_\perp$ 
around $J_2\approx 1.5 J_1$, which would imply a highly non-trivial behavior that 
calls for further research.  Also the role of inter-chain frustration $J_\times$
promises to be an interesting  topic in the future.

\acknowledgments
We thank Masaki Oshikawa, Wei Li, and Zhidan Wang for the useful discussion and/or critical reading of the manuscript.  This research was supported
by the Nachwuchsring of the TU Kaiserslautern,
by the German Research Foundation (DFG) via the Collaborative Research Centers SFB/TR49
and SFB/TR185 (OSCAR),
by the the National Natural Science Foundation of China (Grants No.~11474029 and 11574200),
and by the National Program on Key Research Project (Grants No. 2016YFA0300501).
%
We gratefully acknowledge the computing time granted by the John von Neumann Institute for Computing (NIC)
and provided on the supercomputer JURECA at J\"ulich Supercomputing Centre (JSC).
We also thank the computational resources provided by Physics Laboratory for High Performance Computing~(RUC),
and by Shanghai Supercomputer Center where most of the computations are carried out.

\newpage
\appendix
\section{Bosonization}\label{AppA}
In this appendix we provide additional information for the field theory derivation and the corresponding renormalization group (RG) flow.
For the derivation of the bosonization formulas and the operator product expansion (OPE) it is useful to consider an interacting spinor Fermion model as the underlying physical realization where only the spin channel will be considered in the low-energy limit.\cite{Affleck_1990}
For the half-filled Hubbard model, the charge channel is gapped and the Heisenberg couplings considered in the paper correspond to the spin channel.
The spin currents are expressed as in terms of chiral fermionic field operators
\begin{equation}
J_\kappa^a (z_\kappa)
= :\psi_{\kappa \eta}^{\dag}\frac{\sigma_{\eta \eta'}^a}{2}\psi_{\kappa \eta'}:(z_\kappa),~
\end{equation}
where $\sigma^a$ are the Pauli matrices, the sum over spin components 
$\eta=\uparrow,\downarrow$ is implied, and $\kappa$ denotes the chirality ($\kappa=R/L$ 
or $\kappa=+/-$ respectively).
The chiral complex coordinates are $z_\kappa=-\kappa ix + v\tau$.
The dimerization and staggered magnetization operators are given by \cite{Starykh_2005}
\begin{eqnarray}
\epsilon(z) &\sim & \frac{i}{2}\big[:\psi_{R\eta}^{\dag} \psi_{L\eta}:(z)-:\psi_{L\eta}^{\dag} \psi_{R\eta}:(z) \big],~\nonumber\\
n^a (z) &\sim & \frac{1}{2}\sigma_{\eta\eta'}^a\big[ :\psi_{R\eta}^{\dag}\psi_{L\eta'}:(z)+ 
:\psi_{L\eta}^{\dag}\psi_{R\eta'}:(z)\big],~~~~~
\end{eqnarray}
where $z$ implies a dependence on both chiral variables $z_L$, $z_R$.
The OPEs between $J_\kappa^a$, $\epsilon$, and $n^a$ can be calculated using Wick's theorem \cite{book-cft} and the  two-point correlation function
\begin{equation} 
\langle \psi_{\kappa \eta}(z_\kappa)\psi_{\kappa'\eta'}^\dag(w_{\kappa'})\rangle =
\delta_{\kappa\kappa'}\delta_{\eta\eta'}\frac{\gamma}{z_\kappa-w_\kappa},~ 
\end{equation}
where $\gamma$ depends on the chosen normalization.
Here $\gamma=1/2\pi$.
The required fundamental OPEs are \cite{Starykh_2005}
\begin{eqnarray}\label{fundamental opes}
J^a_\kappa(z_\kappa) \epsilon(w) & =&  i\kappa \frac{\gamma/2}{z_\kappa-w_\kappa} n^a(w),~ \nonumber \\
J^a_\kappa(z_\kappa) n^b(w) &=& i\frac{\gamma/2}{z_\kappa- w_\kappa} 
[\epsilon_{abc} n^c(w) -\kappa\delta_{ab} \epsilon(w)],~\nonumber \\
J^a_\kappa(z_\kappa)J^b_{\kappa'}(w_{\kappa'})&=&\delta_{\kappa\kappa'}\left[ 
\frac{(\gamma^2/2)\delta_{ab}}{(z_\kappa-w_\kappa)^2} +i\epsilon_{abc}\gamma 
\frac{J_\kappa^c(w_\kappa)}{z_\kappa-w_\kappa}\right],~\nonumber\\
\epsilon(z)\epsilon(w)&=& \frac{\gamma^2}{|z-w|}-|z-w| \mathbf{J}_R\cdot \mathbf{J}_L(w),~\nonumber\\
n^a(z)\epsilon(w)&=&-i\gamma |z-w| \left[\frac{J_R^a(w_R)}{z_L- w_L}-\frac{J_L^a(w_L)}{z_R-w_R}\right],~\nonumber\\
n^a(z)n^b(w)&=&|z-w| \left[\hat Q^{ab}(w)+\frac{\gamma^2 \delta_{ab}}{|z-w|^2} 
+i\epsilon_{abc} \gamma \right.\nonumber\\
&\times & \left.\left[\frac{J_R^c(w_R)}{z_L- w_L}+\frac{J_L^c(w_L)}{z_R-w_R}
\right] \right],~
\end{eqnarray}
where $\delta_{ab}$ is the Kronecker $\delta$--function, and $\epsilon_{abc}$ the Levi Civita symbol.
$Q^{ab}$ denotes the zeroth-order contraction between the fermionic fields,
\begin{equation}
\hat Q^{ab} = \frac{1}{2} \sigma_{\eta\eta'}^a \sigma_{\tau\tau'}^b \psi_{R\eta}^\dag \psi_{L\eta'} \psi_{L\tau}^\dag \psi_{R\tau'} .~ 
\end{equation}
As it turns out only the trace of this operator is relevant for the RG flow of spin ladders,
which reads 
\begin{equation}
\hat Q^{aa} = \mathbf{J}_R \cdot \mathbf{J}_L ~
\end{equation}
after freezing out charge degrees of freedom.
The evolution of the bare couplings, is determined from the OPEs of the perturbing operators,  using \cite{Book_Cardy_1996} 
\begin{equation}\label{coupling evolution}
\frac{d\lambda_k}{dl} =(2-d_k)\lambda_k-\frac{\pi}{v}\sum_{i,j} C_{ijk} \lambda_i \lambda_j,~
\end{equation}   
where $d_k$ is the scaling dimension of the operator, $v$ the velocity, and $C_{ijk}$ the coefficient extracted from the OPE
\begin{equation}
O_i (z,\bar z) O_j (w, \bar w ) \sim  \sum_k C_{ijk} \frac{O_k}{(z-w)^{\nu_{k}} (\bar z - \bar w)^{\bar{\nu}_{k}}},
\end{equation}
with $\nu, \bar{\nu}$ the holomorphic and antiholomorphic conformal dimension.
Using Eqs.~(\ref{fundamental opes}) and (\ref{coupling evolution})  we compute the corresponding RG flow.
For the lower part of the phase diagram in Fig.~\ref{figRGPhaseDiagram}, where $J_1 \gg J_2$, $J_\perp$ is assumed, the operators content in Eq.~(\ref{fp1-ops}) can be obtained. 
Using the fundamental OPEs in Eq.~(\ref{fundamental opes}) we arrive at the 
RG equations in Eq.~(\ref{rg-flow-a}).
%
%

%
In the opposite regime, where the coupling $J_2$ is the strongest interaction in the system, $J_2 \gg J_1, J_\perp$, the lattice Hamiltonian can be written as
\begin{eqnarray} \label{hamiltonian2}
H &=&  J_2 \sum_{\bar\alpha=1}^{4} \sum_{j}  \mathbf{S}_{j,\bar\alpha}\cdot \mathbf{S}_{j+1,\bar\alpha} + J_\perp \sum_{j} \mathbf{S}_{j,2}\cdot \mathbf{S}_{j,4} \nonumber \\    
&+& J_1 \sum_{\bar\alpha=1}^{2} \sum_{j}  \mathbf{S}_{j,2\bar\alpha-1}\cdot (\mathbf{S}_{j,2\bar\alpha}  
+ \mathbf{S}_{j+1,2\bar\alpha}). 
\end{eqnarray}
Bosonizing this Hamiltonian gives the operators in Eq.~(\ref{fp2-ops}).
The RG equations in Eq.~(\ref{rg-flow-b}) are derived from the OPEs in Eq.~\eqref{fundamental opes}.
%
%

\begin{figure}[t]
\centering
\includegraphics[width=0.95\columnwidth, clip]{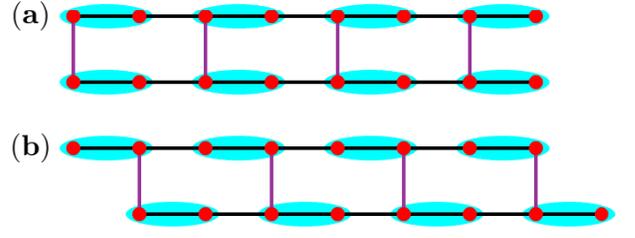}\\
\caption{(a) Configuration of OBC with the corresponding possible columnar pattern of 
singlet dimers (cyan ellipsiods).
(b) Shifted OBC with a staggered dimer pattern.  }\label{FIG-ModelCDvsSD}
\end{figure}
\section{Compatible boundary conditions}\label{AppB}
The CD and SD phases are both doubly degenerate for PBC, since there are two equivalent 
dimer patterns related by symmetry (reflection and translation).
\begin{figure}[t]
\centering
\includegraphics[width=0.95\columnwidth, clip]{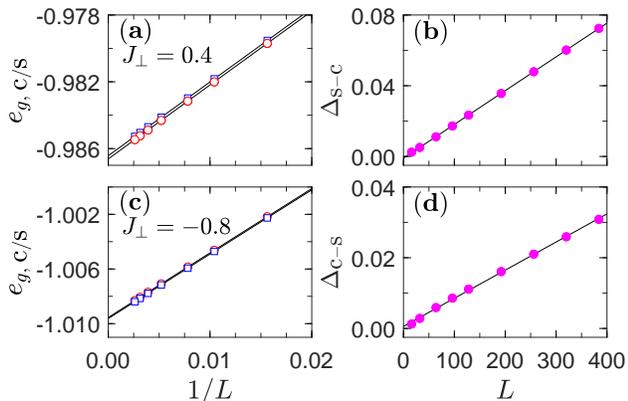}\\
\caption{(a) Ground state energy per-rung $e_{g}=E_{g}/L$ for $J_{\perp}=0.4$ and $J_{2}=1.0$ with OBC and shifted OBC. (b) corresponding total energy difference $\Delta_{\textrm{S}-\textrm{C}} = E_{g}^{\textrm{S}}-E_{g}^{\textrm{C}}$.
Analogous results for $J_{\perp}=-0.8$ and $J_{2}=1.0$ are shown in (c) and (d).}\label{FIG-GSEnergyCDvsSD}
\end{figure}
For OBC, the system will choose the lower energy state with more strong bonds, so
the geometry at edges lifts the degeneracy and fixes the dimer pattern.
In particular, the edges may also affect the relative energy of the SD and CD patterns.
We therefore identify two suitable boundary conditions, which are compatible 
with either the SD or the CD pattern respectively, as shown in Fig.~\ref{FIG-ModelCDvsSD}.
This allows to always identify the lowest possible energetic configuration in the 
simulations.

For the columnar arrangement we use ordinary OBC with an even number of rungs as
shown in Fig.~\ref{FIG-ModelCDvsSD}(a), which prefers the CD phase.
In the staggered pattern, there is a relative shift of the 
dimerization by one site on the two legs, so 
we accordingly also shift the OBC (shifted OBC) as shown in Fig.~\ref{FIG-ModelCDvsSD}(b), 
which prefers the SD phase.  For shifted OBC there is an odd number of rungs but the first
and last rung only contain one spin, so the number of spins remains divisible by 4.
As an example, we now focus on the ground state energies for different 
boundary conditions at $J_{\perp}=0.4$ and $-0.8$ with $J_{2}=1.0$
in Fig.~\ref{FIG-GSEnergyCDvsSD}.
The corresponding ground-state energies 
for both OBC and shifted OBC are shown in Fig.~\ref{FIG-GSEnergyCDvsSD}.
As shown in Fig.~\ref{FIG-GSEnergyCDvsSD}(a), for  $J_{\perp}=0.4$
the ground-state energies per-rung $e_{g,\;\textrm{c}} = -0.986607(8)J_1$ for OBC
and $e_{g,\;\textrm{s}} = -0.986418(6)J_1$ for shifted OBC, 
have a small energy difference of $1.9(1)\times10^{-4}$, independent of $L$. 
The total energy difference $\Delta_{\textrm{S}-\textrm{C}}$ therefore 
increases with $L$ with the corresponding slope shown in Fig.~\ref{FIG-GSEnergyCDvsSD}(b).
This implies that the {\it bulk} energy is affected by the boundaries, so the
correlations of the ground states are different throughout the system (in this case
CD and SD).  Therefore the correct CD pattern can only be obtained with the
corresponding compatible OBC.
Fig.\ref{FIG-GSEnergyCDvsSD}(c,d) show the analogous behavior on the ferromagnetic 
side $J_{\perp}=-0.8$ where shifted OBC must be used to identify the correct 
ground state correlations (SD phase).

\bibliographystyle{apsrev4-1}

\end{document}